\documentclass[11pt]{article}

\usepackage[dvips]{graphicx}
\usepackage{amsmath}
\usepackage{amsthm}
\usepackage{amssymb}

\newcommand{\VV}{2^{\mbox{\scriptsize\boldmath$V$}}}

\newcommand{\sA}{{\mbox{\boldmath$A$}}}
\newcommand{\sB}{{\mbox{\boldmath$B$}}}

\newcommand{\sF}{{\mbox{\boldmath$F$}}}
\newcommand{\sG}{{\mbox{\boldmath$G$}}}

\newcommand{\sV}{{\mbox{\boldmath$V$}}}
\newcommand{\sU}{{\mbox{\boldmath$U$}}}
\newcommand{\sW}{{\mbox{\boldmath$W$}}}
\newcommand{\sX}{{\mbox{\boldmath$X$}}}

\newcommand{\vh}{{\mbox{\boldmath$h$}}}

\newcommand{\as}{{\cal A}}
\newcommand{\AS}{\Gamma}

\newcommand{\QED}{\hfill$\Box$}
\newcommand{\DEF}{\stackrel{\mbox{\scriptsize def}}{=}}

\renewcommand{\l}{\ell}

\newcommand{\halflineskip}{\vspace*{0.5 \baselineskip}}
\renewcommand{\v}[1]{\mbox{\boldmath$ #1 $}}
\newcommand{\sv}[1]{\mbox{\scriptsize \boldmath$ #1 $}}

\newtheoremstyle{test}
     {}
     {}
     {\rm}
     {}
     {\bf}
     {}
     { }
     {}
\theoremstyle{test}

\newtheorem{thm}{Theorem}

\newtheorem{ex}[thm]{Example}
\newtheorem{const}[thm]{Construction}

\newtheorem{rem}[thm]{Remark}

\makeatletter
\def\eqnarray{%
        \stepcounter{equation}%
        \let\@currentlabel=\theequation
        \global\@eqnswtrue\global\@eqcnt\z@
        \tabskip\@centering
        \let\\=\@eqncr
        $$\halign to \displaywidth\bgroup\@eqnsel\hskip\@centering
        $\displaystyle\tabskip\z@{##}$&\global\@eqcnt\@ne
        \hfil$\displaystyle{{}##{}}$\hfil
        &\global\@eqcnt\tw@$\displaystyle\tabskip\z@{##}$\hfil
        \tabskip\@centering&\llap{##}\tabskip\z@\cr}
\makeatother

\title{Optimal Multiple Assignments Based on Integer Programming \\
in Secret Sharing Schemes with General Access Structures\thanks{This
work has been submitted to the IEEE for possible publication. Copyright
may be transferred without notice after which this version may no
longer be accessible.}}
\author{
Mitsugu Iwamoto\thanks{
Graduate School of Information Systems, University of
Electro-Communications, 1-5-1 Chofugaoka, Chofu-shi, Tokyo 182-8585,
Japan. E-mail: {\tt mitsugu@hn.is.uec.ac.jp}} , 
Hirosuke Yamamoto\thanks{
Graduate School of Frontier Science, University of Tokyo, 5-1-5
Kashiwanoha, Kashiwa-shi, Chiba 277-8561, Japan.} , 
and Hirohisa Ogawa\thanks{
C4 technology, Inc., 2-13-17 Kami Ohsaki, Shinagawa-ku,
Tokyo, 141-0021, Japan.}
}
\begin{document}
\maketitle
\begin{abstract}
It is known that for any general access structure, a secret sharing
 scheme (SSS) can be constructed from an $(m,m)$-threshold scheme by 
 using the so-called {\it cumulative map} or from a $(t,m)$-threshold
 SSS by a modified cumulative map. However, such constructed SSSs are
 not efficient generally. In this paper, we propose a new method to
 construct a SSS from a $(t,m)$-threshold scheme for any given general
 access structure. In the proposed method, integer programming
 is used to distribute optimally the shares of $(t,m)$-threshold scheme
 to each participant of the general access structure. From the
 optimality, it can always attain lower coding rate than the cumulative
 maps except the cases that they give the optimal distribution. The same
 method is also applied to construct SSSs for incomplete access
 structures and/or ramp access structures. 
\end{abstract}
\section{Introduction}
A Secret Sharing Scheme \cite{S-cacm,B-afips} (SSS) is a method to
encrypt a secret information $S$ into $n$ pieces called {\it shares}
$V_1,V_2,\ldots,V_n$, each of which has no information of the
secret $S$, but $S$ can be decrypted by collecting several shares. For
example, a $(k,n)$-threshold SSS means that any $k$ out of $n$ shares
can decrypt the secret $S$ although any $k-1$ or less shares do not
leak out any information of $S$. The $(k,n)$-threshold access structure
can be generalized to so-called {\it general access structures} which
consist of the families of {\it qualified sets} and {\it forbidden
sets}. A qualified set is the subset of shares that can decrypt the
secret, but a forbidden set is the subset that does not leak out any
information of $S$.

Generally, the efficiency of a SSS is measured by the entropy of each
share. It is known that for any access structures, the entropies of
secret $S$ and shares $V_i$, $i=1,2,\ldots,n$,  
must satisfy $H(V_i) \ge H(S)$ \cite{KGH-it,CSGV-jc,C-jc}. On the other
hand, in the case of $(k,n)$-threshold SSSs, the optimal SSSs attaining
$H(V_i)=H(S)$ can easily be constructed \cite{S-cacm}. However, it is
hard to derive efficient SSSs for arbitrarily given general access structures although
several construction methods have been proposed.

For example, the {\it monotone circuit construction} \cite{BL-ecrypt88} is
a method to realize a SSS by combining several $(m,m)$-threshold
SSSs. This method is simple but inefficient, and hence, it is extended
to the {\it decomposition construction} \cite{S-it}, which uses  several
decomposed general SSSs. Although the decomposition construction can
attain the optimal coding rates for some special access structures, it
cannot construct an efficient SSS in the case
that the decomposed SSSs cannot be realized efficiently. 
Note that a monotone circuit construction is based on qualified
sets. Hence, as another extension of monotone circuit construction, a
method is proposed to construct a SSS with general access structures
based on qualified sets and $(t,m)$-threshold SSSs \cite{TUM-ieice05}.

On the other hand, for any given general access structure, a SSS can be
constructed from a $(t,m)$-threshold SSS by a multiple assignment map
such that $t$ or more shares of the $(t,m)$-threshold
SSS are assigned to qualified sets but $t-1$ or less shares are assigned
to forbidden sets. The {\it cumulative map} is a simple realization of the
multiple assignment map based on an $(m,m)$-threshold SSS
\cite{ISN-globcom,ISN-ieice,ISN-jc}, and from
the simplicity, it is often used in visual secret sharing schemes for
general access structures \cite{ABSS-ic,KIY-ieice}. However, it is known
that the SSS constructed by the cumulative map is inefficient generally, 
especially in the case that the access structure is a
$(k,n)$-threshold SSS with $k \neq n$. Recently, a {\it modified}
cumulative map based on a $(t,m)$-threshold SSS is
proposed to overcome this defect \cite{T-ieice04}. But, the modified
cumulative map is not always more efficient than the original cumulative
map. 

In this paper, we propose a new construction method that can derive the
optimal multiple assignment map by integer programming. The proposed
construction method is simple and optimal in the sense of multiple
assignment maps. Furthermore, it can also be applied to incomplete and/or 
ramp access structures.  

This paper is organized as follows. In Section 2, we give the
definitions of SSSs and introduce the multiple assignment map. We also
introduce the construction methods of the cumulative map and the modified
cumulative map, and we point out their defects. To overcome such
defects, we propose a new construction method of the optimal multiple
assignment map by integer programming in Section 3. Finally, Sections 4
and 5 are devoted to present the applications of the proposed method to
incomplete or ramp SSSs for general access structures, respectively. 

\section{Preliminaries}
\subsection{Definitions}\label{def.sec}
Throughout this paper, a set of shares and a family of share
sets are represented by bold-face and script letters, respectively. For
sets $\sA$ and $\sB$, we denote a difference set by
$\sA - \sB$, which is defined as $\sA - \sB \DEF \sA
\cap \overline{\sB}$ where $\overline{\sB}$ means the complement of
$\sB$. Furthermore, the cardinality of $\sA$ is represented by $|\sA|$,
and the Cartesian product of $\sA$ and $\sB$ is expressed by $\sA \times
\sB$.

Let $\sV=\{V_1,V_2,\ldots,V_n\}$ be the set of shares, and let $\VV$ be
the family of all subsets of $\sV$. We represent the family of
qualified sets that can decrypt a secret information $S$ and the
family of forbidden sets that cannot gain any information of $S$ by  
$\as_1$ and $\as_0$, respectively.

$\AS=\{\as_1,\as_0\}$ is called an {\em access structure}. For instance,
the access structure of $(k,n)$-threshold SSSs can be
 represented as follows:
\begin{eqnarray}
\label{threshold-1.eq}
\as_1 &=& \{\sA \in \VV: k \le |\sA| \le n\},\\
\label{threshold-2.eq}
\as_0 &=& \{\sA \in \VV: 0 \le |\sA| \le  k-1 \}.
\end{eqnarray}
In SSSs, it obviously holds that $\as_1 \cap \as_0 =
\emptyset$. If it also holds that $\as_1 \cup \as_0 = \VV$, the access
structure is  called {\em complete}. Note
that any access structure must satisfy the following {\em monotonicity}.
\begin{eqnarray}
\label{mono1.eq}
\sA \in \as_1~ \Rightarrow~ \sA' \in \as_1 ~\mbox{\rm for~all}~\sA' \supseteq \sA\\
\label{mono2.eq}
\sA \in \as_0~ \Rightarrow~ \sA' \in \as_0 ~\mbox{\rm for~all}~\sA' \subseteq \sA
\end{eqnarray}
Therefore, we can define the family of {\em minimal} qualified sets and
the family of {\em maximal} forbidden sets as follows:
\begin{eqnarray}
\as^-_1&=&\{\sA\in\as_1:\sA- \{V\} \not\in\as_1~\mbox{\rm for~any}~V\in\sA\},\\
\as^+_0&=&\{\sA\in\as_0:\sA\cup\{V\} \not\in\as_0~\mbox{\rm for~any}~V\in\sV- \sA\}.
\end{eqnarray}

We assume that the secret information $S$ and each share $V_i$ are
random variables, which take values in finite fields ${\mathbb F}_S$ and
${\mathbb F}_{V_i}$, respectively. Then, share set
$\sA=\{V_{i_1},V_{i_2},\ldots,V_{i_u}\}( \subseteq \sV)$, which takes
values in ${\mathbb F}_{\hspace*{-.6mm}\mbox{\boldmath\scriptsize$A$}}
 \DEF
{\mathbb F}_{V_{i_1}} \times {\mathbb F}_{V_{i_2}} \times \cdots \times {\mathbb
F}_{V_{i_u}}$, must satisfy the following conditions: 
\begin{eqnarray}
H(S|\sA)&=&H(S)~~\mbox{\rm if}~~\sA\in\as_0,\\
H(S|\sA)&=&0~~~~~~~\hspace*{.3mm}\mbox{\rm if}~~\sA\in\as_1,
\end{eqnarray}
where $H(S)$ is the entropy of $S$ and $H(S|\sA)$ is the conditional
entropy of $S$ for given $\sA$.

Now, let us define the coding rate of a share $V_i$ as $\rho_i
\DEF H(V_i)/H(S)$, for $i=1,2,\ldots,n$.
 Since each $\rho_i$ may be different in the case of general access
 structures, it is cumbersome to treat each $\rho_i$
 independently. Hence, we consider only the following {\em average}
 coding rate $\tilde \rho$ and {\em worst} coding rate $\rho^*$.
\begin{eqnarray}
\label{ave.eq}
\tilde{\rho}~&\DEF&\frac{1}{n}\sum_{i=1}^n\rho_i,\\
\label{wor.eq}
\rho^\ast    &\DEF&\max_{1 \le i \le n}\rho_i.
\end{eqnarray}

For a given access structure $\AS=\{\as_1,\as_0\}$, we call $V \in \sV$
 a {\em significant} share if there exists a share set $\sA \in \VV$
 such that $\sA \cup \{V\}\in\as_1$ but $\sA \in \as_0$. 
\begin{rem} \label{vac.rem}
Note that a non-significant share plays no roll in the SSS, and hence, 
 $\rho_i=0$ can always be attained for each non-significant share $V_i$
 in any access structure $\AS$. Furthermore, if there exists a
 non-significant share $V_i$ with $\rho_i>0$, the average coding rate
 can be reduced by setting $\rho_i=0$ without changing all the
 significant shares. Hence, we call a non-significant share a {\em
 vacuous} share. On the other hand, we have $\rho_i \ge 1$ for any
 significant share $V_i$ because it must satisfy $H(V_i) \ge H(S)$
 \cite{CSGV-jc,C-jc,KGH-it}. In the following, we assume that every
 share is significant. \QED
\end{rem}

 If a SSS attains $\rho_i=1$ for all
$i$, it is called {\em ideal}. It is known that in the case of 
$(k,n)$-threshold SSSs, the ideal SSS can easily be constructed for any
$k$ and $n$ \cite{S-cacm}. Since $\rho_i \ge 1$,
$i=1,2,\ldots,n$, must hold for any significant share $V_i$ in any
access structures, $\tilde{\rho} = 1$ or $\rho^* =1$ are the necessary
and sufficient conditions for a SSS to be ideal \cite{CSGV-jc}. 
 
\subsection{Multiple Assignment Map}\label{MAS.sec}

Let $\AS=\{\as_1,\as_0\}$ be a given general access structure with share
set $\sV=\{V_1,V_2,\ldots,V_n\}$ and let $\sW_{(t,m)}=\{W^{(t)}_1,
W^{(t)}_2,\ldots,W^{(t)}_m\}$ be the share set of a
$(t,m)$-threshold SSS. We now consider a map
$\varphi_\AS:\{1,2,\ldots,n\} \rightarrow 2^{\mbox{\boldmath
\scriptsize$W$}_{(t,m)}}$, which assigns each participant a
subset of the shares generated by the $(t,m)$-threshold scheme, and a
map $\Phi_\AS:2^{\sv{V}}\rightarrow 2^{\sv{W}_{(t,m)}}$, 
which is defined as $\Phi_{\AS}(\sA) \DEF 
\bigcup_{V_i\in{\mbox{\boldmath\scriptsize$A$}}}\varphi_\AS(i)$
 for a share set $\sA\subseteq\sV$. Then, $\varphi_\AS$ is called a
 {\em multiple assignment map} for the access structure
 $\AS$ if each share $V_i$ is determined by $V_i=\varphi_\AS(i)$ and
 $\Phi_\AS(\sA)$ satisfies the following conditions:
\begin{eqnarray}
\label{mas1.eq}
|\Phi_{\AS}(\sA)|&\ge&
 t~~~~~~~~\hspace*{.3mm}\mbox{\rm if}~\sA\in{\cal A}_1,\\
\label{mas2.eq}
|\Phi_{\AS}(\sA)|&\le&
 t-1~~~\mbox{\rm if}~\sA\in{\cal A}_0,\\
\label{natu-def.eq}
\Phi_\AS(\sV)&=&\sW_{(t,m)}.
\end{eqnarray}
To distinguish $W^{(t)}_j \in \sW_{(t,m)}$ from
the shares $V_i$ of $\AS$,  we call $W^{(t)}_j$ a {\em primitive} share.

Since any $(t,m)$-threshold SSS can easily be constructed as an ideal
SSS \cite{S-cacm,KGH-it}, we assume in this paper that the
$(t,m)$-threshold SSS with
$\sW_{(t,m)}=\{W^{(t)}_1,W^{(t)}_2,\ldots,W^{(t)}_m\}$ is
ideal. Then, the average and worst coding rates defined by 
(\ref{ave.eq}) and (\ref{wor.eq}) become
\begin{eqnarray}
\label{r1.eq}
\tilde{\rho}~&=&\frac{1}{n}\sum_{i=1}^n|\varphi_{\AS}(i)|,\\
\label{r2.eq}
\rho^*&=&\max_{1 \le i \le n}|\varphi_{\AS}(i)|,
\end{eqnarray}
respectively, since it holds that $\rho_i=|\varphi_\AS(i)|$. 

In the case of $t=m$, it is known that the multiple assignment map
$\varphi_\AS$ satisfying (\ref{mas1.eq})--(\ref{natu-def.eq}) can be
realized for any access structures
\cite{ISN-globcom,ISN-ieice,ISN-jc}. Suppose that the access
structure $\AS=\{\as_1,\as_0\}$ has
\begin{eqnarray}\label{F-def.eq}
\as^+_0=\{\sF_1,\sF_2,\ldots,\sF_m\}.
\end{eqnarray}
Note that $m=\left|\as_0^+\right|$. Then, consider the map
$\psi_\AS:\{1,2,\ldots,n\} \rightarrow
2^{\mbox{\boldmath \scriptsize$W$}_{(m,m)}}$ defined by 
\begin{eqnarray}\label{cum.eq}
\psi_{\AS}(i)=\bigcup_{j:V_i\not\in\mbox{\boldmath\scriptsize$F$}_j}\left\{W_j^{(m)}\right\}
\end{eqnarray}
where $\sF_j\in\as_0^+$ and $\sW_{(m,m)}=\{W^{(m)}_1,W^{(m)}_2,\ldots,
W^{(m)}_m\}$
is the set of primitive shares of an $(m,m)$-threshold SSS. The above 
multiple assignment map $\psi_\AS$ is called the {\em
cumulative map}.
\begin{ex} \label{ex1}
Assume that $n=4$ and access structure $\AS_1$ is defined by 
\begin{eqnarray}\label{ex1-1.eq}
\as_1^-&=&\{\{V_1,V_2,V_3\},\{V_1,V_4\},\{V_2,V_4\},\{V_3,V_4\}\},\\
\label{ex1-2.eq}
\as_0^+&=&\{\{V_1,V_2\},\{V_1,V_3\},\{V_2,V_3\},\{V_4\}\}.
\end{eqnarray}
 Then, $m=\left|\as_0^+\right|=4$, and the cumulative map $\psi_{\AS_1}$
 is given from (\ref{cum.eq}) as follows.  
\begin{eqnarray}
V_1&=&\psi_{\AS_1}(1)=\left\{W_3^{(4)},W_4^{(4)}\right\},\\
V_2&=&\psi_{\AS_1}(2)=\left\{W_2^{(4)},W_4^{(4)}\right\},\\
V_3&=&\psi_{\AS_1}(3)=\left\{W_1^{(4)},W_4^{(4)}\right\},\\
V_4&=&\psi_{\AS_1}(4)=\left\{W_1^{(4)},W_2^{(4)},W_3^{(4)}\right\}.
\end{eqnarray}
In this example, it holds that
 $\tilde{\rho}=9/4$ and $\rho^*=3$. \QED
\end{ex}


It is known that the next theorem holds for the cumulative map
$\psi_\AS$. 
\begin{thm}[\cite{SJM-ica}]\label{SJM.thm}
For any multiple assignment map $\varphi_\AS: \{1,2,\ldots,n\} \rightarrow 
2^{\mbox{\boldmath
 \scriptsize$W$}_{(t,m)}}$ with $t=m$, it must hold that
 $|\sW_{(m,m)}|\ge|\as^+_0|$, i.e., $m\ge|{\cal A}_0^+|$. The equality
 holds if and only if $\varphi_\AS(i)$ is equal to the cumulative map
 $\psi_\AS(i)$ defined by (\ref{cum.eq}), where we assume that all
 $\psi_\AS$'s obtained by permutations of $\sF_j$'s in
 (\ref{F-def.eq}) are the same. \QED
\end{thm}

Theorem \ref{SJM.thm} means that, in the case of $t=m$, the cumulative map
$\psi_\AS$ minimizes the number of primitive shares $m$. But, the
minimization of $m$ does not mean the realization of an efficient SSS
generally because it does not minimize the average coding rate
$\tilde\rho$ and/or the worst coding rate $\rho^*$. 

For instance, consider the case that $\AS$ is a $(k,n)$-threshold access
structure with $k \neq n$. If we construct shares $V_i$ by the cumulative map
$\psi$ for this $\AS$, each $V_i$ must consist of $ n-1 \choose k-1$
primitive shares of an $\left( {n \choose k-1 }, {n \choose
k-1}\right)$-threshold SSS because of $|\as^+_0|={n \choose k-1}$. This
means that $\tilde\rho=\rho^*={n-1 \choose k-1}$. But, if we use the
$(k,n)$-threshold SSS itself, we have $\tilde\rho=\rho^*=1$ because each
$V_i$ consists of one primitive share. Hence, the cumulative map is
quite inefficient in the case that
$\AS$ is a $(k,n)$-threshold access structure. In order to
overcome this defect, a {\em modified} cumulative map is proposed in
\cite{T-ieice04} based on $(t,m)$-threshold SSSs. The modified cumulative
map $\psi'_\AS$ is constructed as follows.
\begin{const}[\cite{T-ieice04}]\rm
For a given $\AS=\{\as_0^+,\as_1^-\}$ and a positive integer 
$\displaystyle g\DEF \min_{\mbox{\scriptsize\boldmath$A$}
\in \as_1^-}| \sA|$, 
let ${\cal G}_0\subseteq \as^+_0$ be the family defined by 
\begin{eqnarray}
{\cal G}_0=\{\sG\in{\cal A}^+_0:|\sG|\ge g\}.
\end{eqnarray}
When ${\cal G}_0=\{\sG_1,\sG_2,\ldots,\sG_u\}\neq \emptyset$, 
let $l_j\DEF
|\sG_j|-g+1$ for $j=1,2,\ldots,u$, and ${\l}_j\DEF 
\sum_{p=1}^j l_p$. If ${\cal G}_0 = \emptyset$, let $u=1$ and
 $\ell_1=0$. Then, consider a $(g+{\l}_u,n+{\l}_u)$-threshold SSS
 and the set of primitive shares $\sW_{(g+{\l}_u,n+{\l}_u)}=\{
W^{(g+{\l}_u)}_1, W^{(g+{\l}_u)}_2,\ldots,W^{(g+{\l}_u)}_{n+{\l}_u}\}$. 
Furthermore, let $\sU_j$, $j=1,2,\ldots,u$, be the subset of primitive
 shares defined by 
\begin{eqnarray}
\sU_1&=&\emptyset~~~~~~~~~~
~~~~~~~~~~~~~~~~~~~~~~~~~~~~~~~~~~~
~~~~~~~~\mbox{if}~~~{\cal G}_0=\emptyset,\\
\sU_j&=&\left\{W^{(g+{\l}_u)}_{n+{\l}_{j-1}+1},
W^{(g+{\l}_u)}_{n+{\l}_{j-1}+2}, 
\ldots,W^{(g+{\l}_u)}_{n+{\l}_j}\right\}
~~~~\mbox{if}~~~{\cal G}_0\neq\emptyset,
\end{eqnarray}
where  ${\l}_0=0$. Then, the modified cumulative map $\psi_\AS'$ is
 defined by 
\begin{eqnarray}\label{modi-cm.eq}
\psi_\AS'(i)=\left\{W^{(g+{\l}_u)}_i\right\}\cup\left\{\bigcup_{j:V_i\not
\in\mbox{\scriptsize\boldmath$G$}_j}\sU_j\right\}.
\end{eqnarray}
\QED
\end{const}

In the case where $\AS$ is a $(k,n)$-threshold access structure, it holds
that ${\cal G}_0=\emptyset$ and $\sU_1=\emptyset$, and  hence, it holds
that $\psi_\AS'(i)=\{W^{(k)}_i\}$ for $i=1,2,\ldots,n$ and this scheme
coincides with the ideal $(k,n)$-threshold SSS \cite{T-ieice04}. Therefore,
the modified cumulative map $\psi'_\AS$ is
efficient if $\AS$ is, or is near to, a $(k,n)$-threshold access
structures. Furthermore, it is shown in \cite{T-ieice04} that if the access
structure $\AS$ satisfies
\begin{eqnarray}\label{tochi-cond.eq}
\left|{\cal A}_0^+\right| \ge \frac{(n-g-1) \l_u +n+ 2|{\cal G}_0|}{n-g+1},
\end{eqnarray}
then it holds that for the original cumulative map $\psi_\AS$, 
$\sum_{V_i\in\mbox{\scriptsize\boldmath$V$}}|\psi_\AS'(i)|\le
\sum_{V_i\in\mbox{\scriptsize\boldmath$V$}}|\psi_{\AS}(i)|$, which
means that the average coding rate $\tilde\rho$ of $\psi'_\AS$ is
smaller than or equal to $\psi_\AS$.

But, as shown in the following example, $\psi'_\AS$ is not always more
efficient than $\psi_\AS$ if $\AS$ does not satisfy
(\ref{tochi-cond.eq}). 
\begin{ex} \label{ex2}
Consider the access structure  $\AS_1$ given by
 (\ref{ex1-1.eq}) and (\ref{ex1-2.eq}) in Example \ref{ex1}, which does not
 satisfy (\ref{tochi-cond.eq}). Since we have $g=2$ from
 (\ref{ex1-1.eq}), ${\cal G}_0$ becomes 
${\cal G}_0=\{\{V_1,V_2\},\{V_1,V_3\},\{V_2,V_3\}\} \DEF 
\{\sG_1,\sG_2,\sG_3\}$. Furthermore, since we have that $l_1=l_2=l_3=1$ and
 $\l_3=3$, $\sU_i$'s are determined as 
$\sU_1=\{W^{(5)}_5\},~\sU_2=\{W^{(5)}_6\},~\sU_3=\{W^{(5)}_7\}$ for
 $\sW_{(5,7)}=\{W_1^{(5)},W_2^{(5)},\ldots,W_7^{(5)}\}$. Hence, we can
 check that $\AS_1$ does not satisfy (\ref{tochi-cond.eq}) because of 
$|{\cal A}_0^+|=4$, $n=4$, $g=2$, $\l_u=3$, and $|{\cal G}_0|=3$. 
Finally, we have from (\ref{modi-cm.eq}) that 
\begin{eqnarray}
V_1&=&\psi'_{\AS_1}(1)=\left\{W_1^{(5)},W_7^{(5)}\right\},\\
V_2&=&\psi'_{\AS_1}(2)=\left\{W_2^{(5)},W_6^{(5)}\right\},\\
V_3&=&\psi'_{\AS_1}(3)=\left\{W_3^{(5)},W_5^{(5)}\right\},\\
V_4&=&\psi'_{\AS_1}(4)=\left\{W_4^{(5)},W_5^{(5)},W_6^{(5)},W_7^{(5)}\right\}.
\end{eqnarray}
In this example, the coding rates are given by
 $\tilde{\rho}=5/2$ and $\rho^*=4$, which are larger than the
 coding rates of Example \ref{ex1}, i.e., 
$\tilde\rho=9/4$ and $\rho^*=3$. \QED
\end{ex}

Note that (\ref{tochi-cond.eq}) does not guarantee that the worst coding
rate $\rho^*$ of $\psi'_\AS$ is smaller than $\psi_\AS$. Actually, the
next example shows a case where $\psi_\AS'$
attains  a smaller average coding rate but gives  larger worst coding
rate than $\psi_\AS$. 
\begin{ex}\label{ex3}
Consider the access structure $\AS_2$ given by 
\begin{eqnarray}
\nonumber
{\cal A}^-_1&=&\{\{V_1,V_2,V_3,V_5\},\{V_1,V_2,V_4\},\{V_1,V_3,V_4\},
\{V_1,V_4,V_5\},\\
\label{ex3-1.eq}
&&~~\{V_2,V_3,V_4\},\{V_2,V_4,V_5\},\{V_3,V_4,V_5\}\},\\
\nonumber
{\cal A}^+_0&=&\{\{V_1,V_2,V_3\},\{V_1,V_2,V_5\},\{V_1,V_3,V_5\},
\{V_2,V_3,V_5\},\\
\label{ex3-2.eq}
&&~~\{V_1,V_4\},\{V_2,V_4\},\{V_3,V_4\},\{V_4,V_5\}\}.
\end{eqnarray}
Then, the cumulative map $\psi_{\AS_2}$ is constructed as follows:
\begin{eqnarray}
V_1&=&\psi_{\AS_2}(1)=\left\{W^{(8)}_4,W^{(8)}_6,W^{(8)}_7,W^{(8)}_8\right\},\\
V_2&=&\psi_{\AS_2}(2)=\left\{W^{(8)}_3,W^{(8)}_5,W^{(8)}_7,W^{(8)}_8\right\},\\
V_3&=&\psi_{\AS_2}(3)=\left\{W^{(8)}_2,W^{(8)}_5,W^{(8)}_6,W^{(8)}_8\right\},\\
V_4&=&\psi_{\AS_2}(4)=\left\{W^{(8)}_1,W^{(8)}_2,W^{(8)}_3,W^{(8)}_4\right\},\\
V_5&=&\psi_{\AS_2}(5)=\left\{W^{(8)}_1,W^{(8)}_5,W^{(8)}_6,W^{(8)}_7\right\},
\end{eqnarray}
which attains that $\tilde{\rho}=\rho^*=4$. On the other hand, the
 modified cumulative map $\psi'_{\AS_2}$ is given by
\begin{eqnarray}
V_1&=&\psi'_{\AS_2}(1)=\left\{W^{(7)}_1,W^{(7)}_9\right\},\\
V_2&=&\psi'_{\AS_2}(2)=\left\{W^{(7)}_2,W^{(7)}_8\right\},\\
V_3&=&\psi'_{\AS_2}(3)=\left\{W^{(7)}_3,W^{(7)}_7\right\},\\
V_4&=&\psi'_{\AS_2}(4)=\left\{W^{(7)}_4,W^{(7)}_6,W^{(7)}_7,W^{(7)}_8,W^{(7)}_9\right\},\\
V_5&=&\psi'_{\AS_2}(5)=\left\{W^{(7)}_5,W^{(7)}_6\right\}.
\end{eqnarray}
Observe that the rates of $\psi'_{\AS_2}$ are given by 
$\tilde{\rho}=13/5,~\rho^*=5$. Hence, 
 $\psi_{\AS_2}'$ gives smaller $\tilde{\rho}$ but larger $\rho^*$ than
 $\psi_{\AS_2}$. 
\QED
\end{ex}
As shown in Examples \ref{ex2} and \ref{ex3}, the modified cumulative map
cannot always overcome the defects of the original cumulative
maps. Hence, in the next section, we propose a construction method
of multiple assignment maps that can attain the optimal average or worst
case coding rates based on integer programming. 

\section{Optimal Multiple Assignment Maps}\label{main.sec}
\begin{figure}[bt]
\begin{center}
\includegraphics[width=.6\columnwidth]{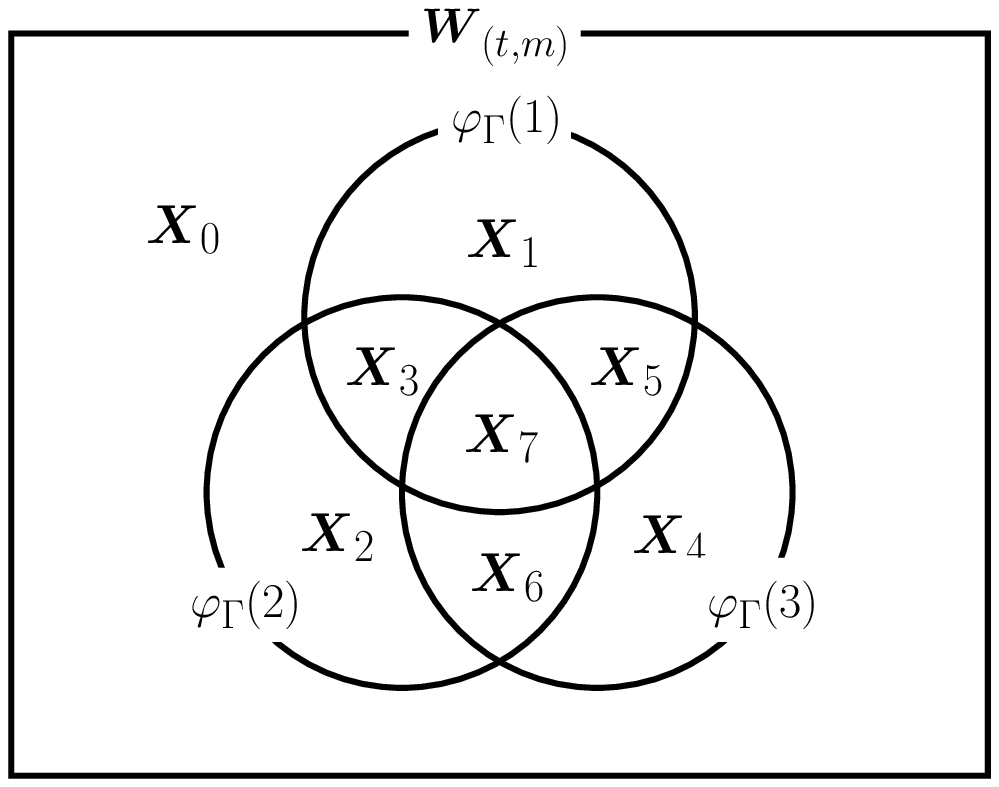}
\end{center}
\caption{Relation between $\varphi_\AS(i)$'s and $\sX_k$'s in the case of $n=3$.}
\label{ven.fig}
\end{figure}

For a multiple assignment map $\varphi_\AS:\{1,2,\ldots,n\} \rightarrow
2^{\mbox{\boldmath\scriptsize$W$}_{(t,m)}}$, a set $\sA \subseteq
\sV$, and $p\in\{0,1,\ldots,2^n-1\}$, let $\sX_p$ be the subset
of $\sW_{(t,m)}$ defined by
\begin{eqnarray}\label{def-X.eq}
\sX_p=\left[\bigcap_{i:b(p)_i=1}\varphi_{\AS}(i)\right]
   \cap\left[\bigcap_{i:b(p)_i=0}\overline{\varphi_{\AS}(i)}\right],
\end{eqnarray}
where $b(p)_i$ is the $i$-th least significant bit in the $n$-bit binary
representation of $p$. For example, in the case of $p=5$ and $n=4$, it
holds that $b(5)_1=b(5)_3=1$, and
(\ref{def-X.eq}) becomes
$\sX_5=\overline{\varphi_\AS(4)}\cap\varphi_\AS(3)\cap
\overline{\varphi_\AS(2)}\cap\varphi_\AS(1)$. Figure \ref{ven.fig} is
the Venn diagram which shows the relation between $\sX_p$'s and
$\varphi_\AS(i)$'s in the case of $n=3$. 
Since $\varphi_\AS$ must satisfy (\ref{natu-def.eq}),  
it must hold that $\bigcap_{i=1}^n\overline{\varphi_\AS(i)}=\emptyset$,
which implies that $\sX_0=\emptyset$. 
Hence, we consider only $\sX_p$ for $p=1,2,\ldots,2^n-1$ in
the following. 

Then, it is easy to check that
$\sX_p$'s satisfy the following equations for an arbitrary $n$ and
$N\DEF 2^n-1$.
\begin{eqnarray}
\label{sa2.eq}
\sX_p \cap \sX_{p'} &=& \emptyset ~~~~\mbox{\rm if}~p \not= p' \\
\label{sa3.eq}
\varphi_\AS(i)&=& \hspace*{-.3cm} \bigcup_{p:b(p)_i=1} \hspace*{-.3cm}\sX_p\\
\label{sa4.eq}
\Phi_\AS(\sA) &=& \bigcup_{V_i \in \mbox{\boldmath\scriptsize$A$}}
\varphi_\AS(i)
= \hspace*{-.5cm} \bigcup_{p:b(p)_i=1
\atop \mbox{\scriptsize for some}~V_i\in\mbox{\boldmath\tiny$A$}} 
\hspace*{-.5cm} \sX_p
\end{eqnarray}

Letting $x_p=|\sX_p|$, the cardinality of $\Phi_\AS(\sA)$ is given by 
\begin{eqnarray}\label{cardinality}
|\Phi_{\AS}(\sA)|=\hspace*{-.3cm}\sum_{p:b(p)_i=1 \atop 
\mbox{\scriptsize for some~} V_i\in\mbox{\boldmath\tiny$A$}} 
\hspace*{-.3cm} x_p,
\end{eqnarray}
from (\ref{sa2.eq}) and (\ref{sa4.eq}).

Now, we describe how to design the optimal multiple assignment map
$\tilde{\varphi}_\AS$ which attains the minimum average coding rate. 
Note that, in order to design the multiple assignment map $\varphi_\AS$
for the set of primitive shares $\sW_{(t,m)}$, 
we have to determine only $x_p$, $p=1,2,\ldots,N$, and $t$, since $m$
can be calculated as $m=\sum_{p=1}^N x_p$ from (\ref{natu-def.eq}) and
(\ref{cardinality}). 

Let $\v{y}\DEF[t,x_1,x_2,\ldots,x_N]$ be the $(N+1)$-dimensional
parameter vector to minimize the average coding rate. 
Furthermore, for an integer $\ell$ and a share set $\sA$, define an
$(N+1)$-dimensional row vector
$\v{a}(\ell;\sA)\DEF[\ell,1(\sA)_1,1(\sA)_2,\ldots,
1(\sA)_N]$ where   
\begin{eqnarray}\label{def-a.eq}
1(\sA)_p=\left\{
\begin{array}{ll}
1&\mbox{\rm if}~b(p)_i=1~\mbox{for some}~V_i\in\v{A}\\
0&\mbox{\rm otherwise.}
\end{array}
\right.
\end{eqnarray}
Then, since (\ref{cardinality}) can be represented by
inner product as $|\Phi_\AS(\v{A})|=\v{a}(0;\sA)\cdot\v{y}^T$ 
where superscript $T$ means the transpose of vector $\v{y}$, the inequalities
in the constraints (\ref{mas1.eq}) and (\ref{mas2.eq}) can be
represented by 
$\v{a}(0;\sA)\cdot\v{y}^T\ge t$, and 
$\v{a}(0;\sA)\cdot\v{y}^T \le t-1$, respectively. Therefore, these
constraints can be expressed as 
\begin{eqnarray}
\label{constraints.eq}
\v{a}(-1;\sA)\cdot\v{y}^T
&\ge&0~~~~\mbox{if}~\sA\in{\cal A}^-_1,\\
-\v{a}(-1;\sA)\cdot\v{y}^T-1
&\ge&0~~~~\mbox{if}~\sA\in{\cal A}^+_0,
\end{eqnarray}
respectively. 
Furthermore, denoting the Hamming weight
in the binary representation of $p$ by $h_p$, it holds from
(\ref{sa3.eq}) that 
\begin{eqnarray}\label{inner}
\sum_{i=1}^n|\varphi_{\AS}(i)|=\sum_{i=1}^n\sum_{p:b(p)_i=1}x_p=
\sum_{p=1}^N h_px_p=\vh\cdot\v{y}^T,
\end{eqnarray}
where $\vh=[h_0,h_1,\ldots,h_N]\in{\mathbb Z}^{N+1}$. Hence, the average
coding rate $\tilde \rho$ in (\ref{r1.eq}) is given by
$(1/n)~\vh\cdot\v{y}^T$ which we want to minimize. 

We note here that $\v{a}(\cdot;\cdot)$ and $\vh$ do not depend on the
 multiple assignment map $\varphi_\AS$, and hence, summarizing
 (\ref{def-a.eq})--(\ref{inner}), we can
formulate the integer programming problem IP${}_{\tilde \rho}(\AS)$ that
minimizes the average coding rate $\tilde{\rho}$ under the constraints
of (\ref{mas1.eq}) and (\ref{mas2.eq}) as follows:

\begin{center}
\begin{tabular}{lr@{}l@{}lcl}
\underline{IP${}_{\tilde \rho}(\AS)$}&&&\\
   minimize & $\vh\cdot\v{y}^T$~~~~~~&     &\\
 subject to & ${\v{a}(-1;\sA)}\cdot\v{y}^T$  &~$\ge$~&~$0$  & for & $\sA\in{\cal A}^-_1$
 \\
            & $-{\v{a}(-1;\sA)}\cdot\v{y}^T$  &~$\ge$~&~$1$ & for & $\sA\in{\cal A}^+_0$
 \\
            & $\v{y}            $     &~$\ge$~&~${\bf 0}$ &     &
\end{tabular}
\end{center}

The optimal multiple assignment map $\tilde\varphi_\AS$  that attains
the minimum average coding rate can be constructed as follows. First,
let $\tilde{\v{y}}=
[\tilde{t},\tilde{x}_1,\tilde{x}_2,\ldots,\tilde{x}_N]$
be the minimizers of the integer programming problem
IP${}_{\tilde \rho}(\AS)$, and we use the $(\tilde t, \tilde
m)$-threshold SSS with primitive shares $\sW_{(\tilde t, \tilde m)}=\{
W_1^{(\tilde t)},W_2^{(\tilde t)},\ldots,W_{\tilde m}^{(\tilde t)}\}$
for secret $S$ 
where $\tilde{m}$ can be calculated from  $\tilde{m}=\sum_{p=1}^N
\tilde{x}_p$. Then, for each $p$, we can assign 
$\tilde x_p$ different primitive shares of $\sW_{(\tilde t, \tilde m)}$
to $\sX_p$ that satisfies $|\sX_p|=\tilde{x}_p$ and
(\ref{sa2.eq}). Finally, the multiple assignment map $\tilde
\varphi_\AS$ is obtained by (\ref{sa3.eq}).

Next, we consider the integer programming problem
IP${}_{\rho^*}(\AS)$ that minimizes the worst coding rate $\rho^*$.
 Let $M$ be the
maximal number of assigned primitive shares among all $V_i$, 
$i=1,2,\ldots,n$. Then, it holds that $|\varphi_\AS(i)|\le M$ for all
$i=1,2,\ldots,n$, and the minimization of $M$  attains 
the optimal worst coding rate. Now, let $\v{z}$ be the
$(N+2)$-dimensional parameter vector defined by 
$\v{z}\DEF[M,t,x_1,x_2,\ldots,x_N]$. Then, it holds that $M=\v{e}
\cdot \v{z}^T$ where $\v{e}$ is the $(N+2)$-dimensional row
vector defined by $\v{e}\DEF[1,0,0,\ldots,0]$. Furthermore,
by defining $\v{b}(\ell,\ell';\sA)\DEF[\ell,\ell',1(\sA)_1,1(\sA)_2,
\ldots,1(\sA)_N]$ where $1(\sA)_p$ is defined by (\ref{def-a.eq}), the
number of
primitive shares assigned to a share set $\sA\subseteq\sV$ can be
expressed as $\v{b}(0,0;\sA)\cdot \v{z}^T$. 
Hence, in the same way as IP${}_{\tilde\rho}(\AS)$, the integer
programming problem IP${}_{\rho^*}(\AS)$ that minimizes the worst coding
rate $\rho^*$ can be formulated as follows: 
\begin{center}
\begin{tabular}{lr@{}l@{}lcl}
\underline{IP${}_{\rho^*}(\AS)$}&&\\
   minimize &  $\v{e}\cdot \v{z}^T$ &     &\\
 subject to & ${\v{b}(0,-1;\sA)}\cdot\v{z}^T$  &~$\ge$~~& $0$  & \mbox{\rm for} &
 $\sA\in{\cal A}^-_1$ \\
            & $-{\v{b}(0,-1;\sA)}\cdot\v{z}^T$  &~$\ge$~~& $1$ & \mbox{\rm for} &
 $\sA\in{\cal A}^+_0$ \\
            & $-{\v{b}(-1,0;\{V\})}\cdot\v{z}^T$&~$\ge$~~& $0$ &   \mbox{\rm for} &
 $V\in\sV$ \\
            & $\v{z}            $     &~$\ge$~~&${\bf 0}$ &     &
\end{tabular}
\end{center}

The multiple assignment map $\varphi^*_\AS$ attaining the minimum
$\rho^*$ can also be constructed from the obtained minimizer in the same
way as the construction of $\tilde\varphi_\AS$.

\begin{rem}
Actually, in SSSs, we can assume without loss of generality that
 $x_N=0$, i.e., $\sX_N=\bigcap_{i=1}^n\varphi_\AS(i)=\emptyset$ because
 it is not necessary to consider the set of primitive shares commonly
 contained in every share. Hence, the vectors in integer
 programming problems {\rm IP}${}_{\tilde\rho}(\AS)$ and {\rm
 IP}${}_{\rho^*}(\AS)$ can be reduced to $N$-dimensional and
 $(N+1)$-dimensional vectors, respectively. However, $x_N=0$ does not
 hold generally in the case of ramp SS schemes, which is described in
 Remark \ref{x_n.rem} in Section \ref{OMAMR.sec}.\QED
\end{rem}

\begin{ex}
For the access structure $\AS_1$ defined by
 (\ref{ex1-1.eq}) and (\ref{ex1-2.eq}) in Example \ref{ex1}, the integer
 programming problem {\rm IP}${}_{\tilde \rho}(\AS_1)$ can be formulated as
 follows:
\begin{center}
\begin{tabular}{@{}ll@{}}
\underline{{\rm IP}${}_{\tilde \rho}(\AS_1)$} & \\
{\rm minimize}&$x_1+x_2+2x_3+x_4+2x_5+2x_6+3x_7+x_8+2x_9+2x_{10}$\\
               & \hspace*{8cm}$+3x_{11}+2x_{12}+3x_{13}+3x_{14}$ \\
{\rm subject to} &$-t+x_1 + x_2 + x_3 + x_4 + x_5 + x_6 + x_7 + x_9 $
\hspace*{3.5cm}\\&
\begin{tabular}{r@{}l@{}l}
$+ x_{10} + x_{11} + x_{12}+ x_{13} + x_{14}$  ~& $\ge$&~$0$ \\
$-t+  x_1  + x_3  + x_5 + x_7 + x_8 + x_9 + x_{10} + x_{11} + x_{12} + x_{13} + x_{14}$  ~&$\ge$&~$0$ \\
$  -t+ x_2 + x_3 + x_6 + x_7 + x_8 + x_9 + x_{10} + x_{11} + x_{12} + x_{13} + x_{14}$  ~&$\ge$&~$0$ \\
$-t+  x_4 + x_5 + x_6 + x_7 + x_8 + x_9 + x_{10} + x_{11} + x_{12} + x_{13} + x_{14}$  ~&$\ge$&~$0$ \\ 
$ t- x_1 - x_2 - x_3 - x_5 - x_6 - x_7  - x_9 - x_{10} - x_{11} - x_{13} - x_{14}$  ~&$\ge$&~$1$ \\
$ t- x_1 - x_3 - x_4 - x_5 - x_6 - x_7 - x_9 - x_{11} - x_{12} - x_{13} - x_{14}$  ~&$\ge$&~$1$ \\
$ t- x_2 - x_3 - x_4 - x_5 - x_6 - x_7 - x_{10} - x_{11} - x_{12} - x_{13} - x_{14}$  ~&$\ge$&~$1$ \\
$ t- x_8 - x_9 - x_{10} - x_{11} - x_{12} - x_{13} - x_{14}$ ~&$\ge$&~$1$ \\
\hfill$x_p$ ~&$\ge$&~ $0,p=1,2,\ldots,14$\\
\end{tabular}
\end{tabular}
\end{center}
\halflineskip

By solving the above {\rm IP}${}_{\tilde \rho} (\AS_1)$, we obtain that 
the value of the objective function is $5$, which is attained by the
 following minimizers: 
\begin{eqnarray}\label{so.eq}
\tilde{t}=3, ~\tilde{x}_1&=&\tilde{x}_2=\tilde{x}_4=1,~\tilde{x}_8=2,~
\tilde x_i=0~~~{\rm for}~i=3,5,6,7,9,10,\ldots,14,
\end{eqnarray}
Hence, $\tilde m$ is given by $\tilde{m}=\sum_{p=1}^{14}\tilde x_p=5$,
 and $\sX_p$'s  become 
\begin{eqnarray}
\sX_1 &=& \left\{W_1^{(3)}\right\},~~\sX_2=\left\{W_2^{(3)}\right\},~~
\sX_4 = \left\{W_3^{(3)}\right\},~~\sX_8=\left\{W_4^{(3)},W_5^{(3)}\right\},
\end{eqnarray}
where $\sW_{(3,5)} =\{W_1^{(3)},W_2^{(3)},\ldots, W_5^{(3)}\}$. 
Finally, from (\ref{sa3.eq}), $\tilde{\varphi}_{\AS_1}$ is constructed as 
\begin{eqnarray}
V_1&=&\tilde \varphi_{\AS_1}(1)=\left\{W_1^{(3)}\right\},\\
V_2&=&\tilde \varphi_{\AS_1}(2)=\left\{W_2^{(3)}\right\},\\
V_3&=&\tilde \varphi_{\AS_1}(3)=\left\{W_3^{(3)}\right\},\\
V_4&=&\tilde \varphi_{\AS_1}(4)=\left\{W_4^{(3)},W_5^{(3)}\right\}.
\end{eqnarray}
In this case, we have that $\tilde{\rho}=5/4$ and ${\rho}^*=2$. The
 integer programming problem $\rm IP_{\hspace*{-.5mm}\rho^*}({\AS_1})$
 derives the same solutions as (\ref{so.eq}),
 and hence, it holds that $\tilde{\varphi}_{\AS_1}= \varphi_{\AS_1}^*$
 in this example. Recall that the cumulative map $\psi_{\AS_1}$ attains
 the coding rates $\tilde{\rho}=9/4$ and $\rho^*=3$, and the modified
 cumulative map $\psi'_{\AS_1}$ attains $\tilde{\rho}=5/2$ and
 $\rho^*=4$. Hence, $\varphi_{\AS_1}$ can attain smaller coding rates
 compared with $\psi_{\AS_1}$ and $\psi'_{\AS_1}$. \QED
\end{ex}

\begin{ex}
For the access structure $\AS_2$ defined by 
 (\ref{ex3-1.eq}) and (\ref{ex3-2.eq}) in Example \ref{ex3}, we can obtain
 the following multiple assignment map by solving the integer
 programming problem ${\rm IP}_{\hspace*{-.5mm}\tilde\rho}(\AS_2)$.
\begin{eqnarray}
\label{AS2-1.eq}
V_1&=&\tilde\varphi_{\AS_2}(1) = \left\{W_1^{(4)}\right\}, \\
V_2&=&\tilde\varphi_{\AS_2}(2) = \left\{W_2^{(4)}\right\}, \\
V_3&=&\tilde\varphi_{\AS_2}(3) = \left\{W_3^{(4)}\right\}, \\
V_4&=&\tilde\varphi_{\AS_2}(4) = \left\{W_4^{(4)},W_5^{(4)}\right\}, \\
\label{AS2-2.eq}
V_5&=&\tilde\varphi_{\AS_2}(5) = \left\{W_6^{(4)}\right\}, 
\end{eqnarray}
where $W_i^{(4)} \in \sW_{(4,6)}$. Then, it holds that
 $\tilde\rho=6/5$ and $\rho^* = 2$. Furthermore, it holds that
 $\tilde{\varphi}_{\AS_2}=\varphi^*_{\AS_2}$ in this access structure. 
Recall again that the cumulative map $\psi_{\AS_2}$ attains
 the coding rates $\tilde{\rho}=\rho^*=4$, and the modified
 cumulative map $\psi'_{\AS_2}$ attains $\tilde{\rho}=13/5$ and
 $\rho^*=5$. Hence, $\tilde{\varphi}_{\AS_2}$ is more efficient than
 $\psi_{\AS_2}$ and $\psi'_{\AS_2}$. 
\QED
\end{ex}

Since any access structure can be realized by the cumulative map (and
the modified cumulative map), there exists at least one multiple
assignment map for any access structure. Therefore, the next theorem
holds obviously. 
\begin{thm}
For any access structure $\AS$ that satisfies monotonicity
 (\ref{mono1.eq}) and (\ref{mono2.eq}), the integer programming problems
 {\rm IP}${}_{\tilde\rho}(\AS)$ and {\rm IP}${}_{\rho^*}(\AS)$ always
 have at least one feasible solution, and hence, there exists the
 optimal multiple assignment map. \QED
\end{thm}

We note that the integer programming problems are NP-hard, and
hence, the proposed algorithms may take much time in
solving for large $n$ $(=|\sV|)$. But, in the case that $n$ is not large,
the solution is obtained quickly. For instance, in the case of
{\rm IP}$_{\rho}(\AS_3)$ in Example \ref{tochi.ex} with $n=6$, it can be solved
within $0.1$ seconds by a notebook computer. 
\begin{ex}\label{tochi.ex}
Consider the following access structure $\AS_3$:
\begin{eqnarray}
\nonumber
\as^-_1&=&\{\{V_1,V_3,V_4,V_5\},\{V_1,V_3,V_5,V_6\},\{V_1,V_4,V_5,V_6\},\{V_3,V_4,V_5,V_6\},
             \{V_1,V_2,V_3\},\{V_1,V_2,V_5\},\\
\nonumber
&& ~~\{V_1,V_2,V_6\},\{V_2,V_3,V_4\},\{V_2,V_3,V_5\},\{V_2,V_3,V_6\},\{V_2,V_4,V_5\},\{V_2,V_4,V_6\},\{V_2,V_5,V_6\}\},\\
\label{tochi1.eq}\\
\nonumber
\as^+_0&=&\{\{V_1,V_3,V_4,V_6\},\{V_1,V_2,V_4\},\{V_1,V_3,V_5\},\{V_1,V_4,V_5\},\{V_1,V_5,V_6\},\{V_3,V_4,V_5\},\\
&&~~\{V_3,V_5,V_6\},\{V_4,V_5,V_6\},\{V_2,V_3\},\{V_2,V_5\},\{V_2,V_6\}\}.
\label{tochi2.eq}
\end{eqnarray}
Then, we obtain the following multiple assignment map by solving
 {\rm IP}${}_{\tilde\rho}\left(\AS_3\right)$.
\begin{eqnarray}
V_1 &=& \tilde\varphi_{\AS_3}(1)=\left\{W^{(6)}_1,W^{(6)}_2\right\},\\
V_2 &=& \tilde\varphi_{\AS_3}(2)=\left\{W^{(6)}_1,W^{(6)}_3,W^{(6)}_4,W^{(6)}_5\right\},\\
V_3 &=& \tilde\varphi_{\AS_3}(3)=\left\{W^{(6)}_6\right\},\\
V_4 &=& \tilde\varphi_{\AS_3}(4)=\left\{W^{(6)}_2,W^{(6)}_5\right\},\\
V_5 &=& \tilde\varphi_{\AS_3}(5)=\left\{W^{(6)}_3,W^{(6)}_7\right\},\\
V_6 &=& \tilde\varphi_{\AS_3}(6)=\left\{W^{(6)}_8\right\},
\end{eqnarray}
where $W^{(6)}_i\in\sW_{(6,8)}$. $\tilde\varphi_{\AS_3}$ attains
 that $\tilde\rho=2$ and $\rho^*=4$. On the other hand, the
 cumulative map for the access structure $\AS_3$ are given by 
\begin{eqnarray}
V_1 &=& \psi_{\AS_3}(1)=\left\{W^{(11)}_6,W^{(11)}_7,W^{(11)}_8,W^{(11)}_9,W^{(11)}_{10},W^{(11)}_{11}\right\},\\
V_2 &=& \psi_{\AS_3}(2)=\left\{W^{(11)}_1,W^{(11)}_3,W^{(11)}_4,W^{(11)}_5,W^{(11)}_6   ,W^{(11)}_7,  W^{(11)}_8\right\},\\
V_3 &=& \psi_{\AS_3}(3)=\left\{W^{(11)}_2,W^{(11)}_4,W^{(11)}_5,W^{(11)}_8,W^{(11)}_{10},W^{(11)}_{11}\right\},\\
V_4 &=& \psi_{\AS_3}(4)=\left\{W^{(11)}_3,W^{(11)}_5,W^{(11)}_7,W^{(11)}_9,W^{(11)}_{10},W^{(11)}_{11}\right\},\\
V_5 &=& \psi_{\AS_3}(5)=\left\{W^{(11)}_1,W^{(11)}_2,W^{(11)}_9,W^{(11)}_{11}\right\}, \\
V_6 &=& \psi_{\AS_3}(6)=\left\{W^{(11)}_2,W^{(11)}_3,W^{(11)}_4,W^{(11)}_6,W^{(11)}_9,W^{(11)}_{10}\right\},
\end{eqnarray}
where $W^{(11)}_i\in\sW_{(11,11)}$. $\psi_{\AS_3}$ has
 $\tilde\rho=35/6$ and  $\rho^*=7$. Furthermore, the modified
 cumulative map for $\AS_3$ requires $(12,15)$-threshold SSS and has
 $\tilde\rho=5$ and $\rho^*=9$. \QED
\end{ex}

Next, we clarify what kind of access structure can be realized as an
ideal SSS by the multiple assignment map. 
\begin{thm}\label{ideal-condition.thm}
For an access structure $\AS$, the SSS constructed by the optimal
 multiple assignment map is ideal,
 i.e., $\rho_i=1$ for all $i$, if and only if $\as_1^-$ of $\AS$ can be
 represented by
\begin{eqnarray}
\label{ns.eq}
 {\cal A}^-_1&=&\bigcup_{\forall\{j_1,j_2,\ldots,j_t\} \atop
\subseteq \{1,2,\ldots,m\}}
\left\{ \sA_{j_1} \times \sA_{j_2} \times \cdots \times \sA_{j_t} \right\},
\end{eqnarray}
where $t$ is a positive integer and
 $\{\sA_1,\sA_2,\ldots,\sA_m\}$ is a partition of $\sV$ which satisfies 
\begin{eqnarray}
\label{cup.eq}
 \bigcup_{j=1}^m \sA_j &=& \sV,\\
\label{*.eq}
\sA_j &\neq& \emptyset~~~~~\mbox{\rm for}~j=1,2,\ldots,m, \\
\label{dis.eq}
 \sA_j \cap \sA_{j'}&=&\emptyset~~~~~\mbox{\rm if}~j \not= j'.
\end{eqnarray}
\QED
\end{thm}

{\em Proof of Theorem  \ref{ideal-condition.thm}:} If there exists a 
partition $\{\sA_1,\sA_2,\ldots,\sA_m\}$ satisfying
(\ref{ns.eq})--(\ref{dis.eq}) for the access structure $\AS$, the ideal
SSS can be obtained by letting
\begin{eqnarray}\label{representation.eq}
\varphi_\AS(i)=W^{(t)}_j~~~\mbox{\rm if}~V_i \in \sA_j
\end{eqnarray}
for each $i=1,2,\ldots,n$. Next, we show the necessity of
(\ref{ns.eq})--(\ref{dis.eq}). Suppose that a certain $\varphi_\AS(i)$
attains $\rho_i=1$ for all $i$. Then, define each $\v{A}_j$ as 
\begin{eqnarray}\label{ideal.eq}
\v{A}_j\DEF \Phi_\AS^{-1}\left(\left\{W_j^{(t)}\right\}\right),~j=1,2,\ldots,m,
\end{eqnarray}
for $j=1,2,\ldots,m$ where $\Phi_\AS^{-1}: 
2^{\mbox{\scriptsize\boldmath$W$}_{(t,m)}} 
\rightarrow \VV$ is the inverse map of
$\Phi_\AS(\v{A})\DEF\sum_{i:V_i\in\sv{A}}\varphi_\AS(i)$. Then, it is easy to
see that $\sA_j$'s satisfy (\ref{ns.eq}), (\ref{cup.eq}) and
(\ref{*.eq}). Next, we prove that $\v{A}_j$'s defined by
(\ref{ideal.eq}) satisfy (\ref{dis.eq}). Assume that there exist $\sA_j$
and $\sA_{j'}$, $j \neq j'$, not satisfying (\ref{dis.eq}). 
Then, there exists a share
$V_i\in\sA_j\cap\sA_{j'}$. This means that $\varphi_{\AS}(i)
\supseteq \{W^{(t)}_j,W^{(t)}_{j'}\}$, which contradicts
$\rho_i=|\varphi_\AS(i)|=1$. Hence, $\{\sA_1,\sA_2,\ldots,\sA_m\}$ must
be a partition of $\sV$ satisfying (\ref{ns.eq})--(\ref{dis.eq}). \QED

\halflineskip
In the case of $t=2$, it is known that an access structure $\AS$ can be
realized by an ideal SSS if and only if $\AS$ can be represented by a
complete multipartite graph \cite{BSSV-jc}. We note that this condition
coincides with (\ref{ns.eq})--(\ref{dis.eq}) in this case. Furthermore,
in the case that  $|\sA_j|=1$ for $j=1,2,\ldots,m$, the access structure
coincides with the  $(t,m)$-threshold access structure. Hence, if $\AS$ is the $(k,n)$-threshold access structure, the multiple assignment maps obtained from the integer programming problems {\rm IP}$_{\tilde\rho}(\AS)$ and {\rm
 IP}$_{\rho^*}(\AS)$ obviously satisfy that
 $|\tilde\varphi_{\AS}(i)|=|\varphi^*_{\AS}(i)|=1$ for all $i$. 

We note that any access structures not satisfying
(\ref{ns.eq})--(\ref{dis.eq}) must have $\tilde{\rho}>1$ and $\rho^*\ge
2$ if the multiple assignment map is used. But, an access structure not
satisfying (\ref{ns.eq})--(\ref{dis.eq}) might be realized as an ideal
SSS if we use another construction method. For example, refer \cite{S-it}. 

In this paper, we assume that every share is significant. But, if there
exist vacuous shares in the access structure $\AS$, it is cumbersome to
check whether each share is significant or vacuous. From Remark
\ref{vac.rem}, the optimal multiple assignment map $\tilde\varphi_\AS$
attaining the minimum average coding rate  must satisfy that
$|\tilde\varphi_\AS(i)|=0$ for any vacuous share $V_i$. On the other
hand, it clearly holds that $|\varphi_\AS(i)|\ge 1$ for every
significant share $V_i$ since $\rho_i\ge 1$ holds for any significant
share. Hence, by solving the integer programming problem
IP$_{\tilde\rho}(\AS)$, we can also know whether a share is significant or
vacuous. 

\section{Multiple Assignment Maps for Incomplete Access Structures}
\label{IAS.sec}

In the previous sections, we considered how to construct a SSS for a
complete general access structure $\AS=\{\as_1,\as_0\}$. 
But in practice, it may be cumbersome 
to specify whether each subset of $\sV$ is a qualified set or a
forbidden set because the number of the subsets is $2^n$. Hence, a method
is proposed in \cite{ISN-jc} to construct a SSS for the case such that
some subsets of $\sV$ are not specified as qualified nor forbidden sets. 

\begin{thm}[\cite{ISN-jc}] Let $\AS^\sharp=\{\as^\sharp_1, \as^\sharp_0\}$ 
be an incomplete access structure, which has $\as_1^\sharp\cup
\as^\sharp_1\neq \VV$. 
Then, there exists a complete access structure $\AS=\{\as_1,\as_0\}$
 such that 
\begin{eqnarray}
\label{ab1.eq}
\as^\sharp_1 &\subseteq& \as_1,\\
\label{ab2.eq}
\as^\sharp_0 &\subseteq& \as_0,
\end{eqnarray}
if and only if it holds that for any $\sA \in\as^\sharp_1$ and
 $\sB \in\as^\sharp_0$, 
\begin{eqnarray}\label{incomp.eq}
 \sA \nsubseteq \sB.
\end{eqnarray}
\QED
\end{thm}

In case that (\ref{incomp.eq}) is satisfied, the SSS satisfying the incomplete
access structure $\AS^\sharp=\{\as^\sharp_1,\as^\sharp_0\}$ can be realized by
applying the cumulative map to the complete access structure
$\AS=\{\as_1,\as_0\}$. In fact, for the access structure
$\AS^\sharp=\{\as^\sharp_1,\as^\sharp_0\}$, a SSS  is constructed in
\cite{ISN-jc} by a cumulative map $\psi_{\AS^\sharp}(i) =
\bigcup_{j:V_i\not\in{\mbox{\boldmath\scriptsize$F$}}_j}\{W^{(t)}_j \}$ for
$\as_0^{\sharp+}=\left\{{\sF_1},{\sF_2},\ldots,{\sF_m}\right\}$.
This construction corresponds to the case that 
\begin{eqnarray}
\label{incomplete-cm.eq}
\as_0^+=\as_0^{\sharp+} ~\mbox{\rm and}~\as_1=\VV-\as_0.
\end{eqnarray}

However, $\psi_{\AS^\sharp}$ is not efficient generally because
$\psi_{\AS^\sharp}$ is a cumulative map, which is inefficient
as described in Section \ref{MAS.sec}. Furthermore, even if the
cumulative map can attain the optimal coding rates for the access
structure given by (\ref{incomplete-cm.eq}), the access structure may
not be optimal among all the complete access structures
$\AS=\{\as_1,\as_0\}$ satisfying (\ref{ab1.eq}) and (\ref{ab2.eq}) for
given $\AS^\sharp=\{\as_1^\sharp,\as_0^\sharp\}$. 

In our construction based on integer programming, the optimal multiple
assignment map for the incomplete access structure
$\AS^\sharp=\{\as^{\sharp-}_1,\as^{\sharp+}_0\}$ can easily be obtained
by applying IP${}_{\tilde\rho}(\Gamma)$ or IP${}_{\rho^*}(\Gamma)$
directly to $\AS^\sharp$.

\begin{ex}\rm Let us consider the following access structure $\AS^\sharp_3=
\{\as^\sharp_1,\as^\sharp_0\}$:
\begin{eqnarray}
\label{tochi-a.ex}
\as^\sharp_1&=&\{\{V_1,V_4,V_5,V_6\},\{V_1,V_2,V_5\},\{V_1,V_2,V_6\},\{V_2,V_3,V_6\},\{V_2,V_4,V_6\}\},\\
\label{tochi-b.ex}
\as^\sharp_0&=&\{\{V_1,V_3,V_4,V_6\},\{V_1,V_3,V_5\},\{V_1,V_5,V_6\},\{V_3,V_4,V_5\},\{V_4,V_5,V_6\},\{V_2,V_5\}\},
\end{eqnarray}
Note that $\as^\sharp_1$ and $\as^\sharp_0$ satisfy 
 $\as^\sharp_1 \subseteq \as_1^-$ and $\as^\sharp_0 \subseteq \as_0^+$
 for $\AS_3=\{\as_1,\as_0\}$, which is 
 defined by (\ref{tochi1.eq}) and (\ref{tochi2.eq}) in Example
 \ref{tochi.ex}. Then, by solving ${\rm IP}_{\tilde\rho}(\AS^\sharp_3)$,
 we obtain the following multiple assignment map.
\begin{eqnarray}
V_1 &=& \tilde{\varphi}_{\AS_3^\sharp}(1) = \left\{ W_1^{(4)}\right\},\\
V_2 &=& \tilde{\varphi}_{\AS_3^\sharp}(2) = \left\{ W_2^{(4)},W_3^{(4)}\right\},\\
V_3 &=& \tilde{\varphi}_{\AS_3^\sharp}(3) = \left\{ W_4^{(4)}\right\},\\
V_4 &=& \tilde{\varphi}_{\AS_3^\sharp}(4) = \left\{ W_4^{(4)}\right\},\\
V_5 &=& \tilde{\varphi}_{\AS_3^\sharp}(5) = \left\{ W_5^{(4)}\right\},\\
V_6 &=& \tilde{\varphi}_{\AS_3^\sharp}(6) = \left\{ W_6^{(4)}\right\},
\end{eqnarray}
where $W_i^{(4)}\in\sW_{(4,6)}$, and it holds that $\tilde\rho = 7/6$ 
and $\rho^*=2$. If we apply the cumulative map to $\AS_3^\sharp$,
 $\psi_{\AS_3^\sharp}$ is constructed from the $(6,6)$-threshold
 scheme, and it has $\tilde\rho=3$ and $\rho^*=5$. \QED
\end{ex}

Similarly to the complete SSS, vacuous shares $V_i$ in
$\AS^\sharp=\{\as^\sharp_1,\as^\sharp_0\}$ can be detected  by checking 
$|\varphi_{\AS^\sharp}(i)|=0$ for the solution of the 
IP$_{\tilde\rho}(\AS^\sharp)$. 

\section{Ramp SSSs with General Access Structures}

The coding rate $\rho_i$ must satisfy $\rho_i \ge 1$ for any significant
share $V_i$ in the case that the access structure consists of $\as_1$
and $\as_0$, i.e., every subset $\sA
\subseteq \sV$ is classified into either qualified sets or  forbidden
sets. But, in the case of ramp access structures such that some subsets
of $\sV$ are allowed to have intermediate properties between the qualified
and forbidden sets, it is possible to decrease the coding rate
$\rho_i$ to less than 1. The SSSs having the ramp access structure are called
{\em ramp schemes} \cite{BM-crypto85,HY-ieice}. In this section, we
treat the construction of ramp SSSs based on the multiple assignment
maps. We consider only the minimum average coding rate in this section. But, for the
minimum worst coding rate, integer programming can be formulated in
a similar way. 
\subsection{Preliminaries for Ramp Schemes}
First, let us review the definition of ramp SSSs. Suppose that $L+1$
families $\as_j \subseteq \VV$, $j=0,1,\ldots,L$, satisfy the following.
\begin{eqnarray}\label{ramp-def.eq}
H(S|\sA)=\frac{L-j}{L}H(S),~~\mbox{\rm for~any~} \sA \in\as_j
\end{eqnarray}
Equation (\ref{ramp-def.eq}) implies that the secret $S$ leaks out
from a set $\sA \in \as_j$ with the amount of
$(j/L)H(S)$. Especially, $S$ can be decrypted completely from any
$\sA \in \as_L$, and any $\sA \in \as_0$ leaks out no information of
$S$. Note that, in the case of $L=1$, the ramp SSS reduces to the SSS
treated in Sections 2--4, and hence, the ramp SSS can be considered as
an extension of the ordinal SSS. To distinguish the ordinal SSSs from ramp
SSSs, the ordinal SSSs are called the {\em perfect} SSSs. We
call $\AS^R = \{ \as_0,\as_1,\ldots,\as_L\}$ the access
structure of the ramp SSS with $L+1$ levels. Without loss of 
generality, we can assume that $\bigcup_{j=0}^L\as_j =\VV$ and $\as_j
\cap \as_{j'} = \emptyset$ for $j\neq j'$, although incomplete access
structures with $\bigcup_{j=0}^L\as_j \neq \VV$ can be treated in the
same way as in Section \ref{IAS.sec}.

For example, the access structure of $(k,L,n)$-threshold ramp SSS
\cite{HY-ieice,BM-crypto85} is defined as follows:
\begin{eqnarray}
\label{Sham.eq1}
\as_0&=&\{\sA\in\VV : 0 \le |\sA| \le k-L \},\\
\as_j&=&\{\sA\in\VV : |\sA|=k-L+j\},~~\mbox{\rm for~}1 \le j \le L-1,\\
\label{Sham.eq3}
\as_L&=&\{\sA\in\VV : k \le |\sA| \le n\}.
\end{eqnarray}

In  ramp SSSs, a significant share can also be defined in the same
way as the perfect SSSs shown in Section \ref{def.sec}. A share
$V_i\in \sV$ is called {\em significant} if there exists a share set
$\sA\in\VV$ such that $\sA\cup\{V_i\}\in\as_j$ and $\sA\in\as_{j'}$ with
$j>j'$. Then, a non-significant share $V_{i'}$ satisfies that
$\sA\cup\{V_{i'}\}\in\as_j$ for any share set $\sA\in\as_j$,
$j=0,1,\ldots,L$. Furthermore, if a non-significant share $V_{i'}$
satisfies $\{V_{i'}\}\in\as_0$, $V_{i'}$ plays no roll in the ramp SSS,
and hence, we call $V_{i'}$ a {\em vacuous} share. However, there exists
a ramp scheme such that $\as_0=\emptyset$ and a non-significant share
satisfy $\{V_i\}\in\as_j$ for some $j \ge 1$. This case implies that
$H(V_{i'})\ge H(S)/L$, and $H(V_{i'}|V)=0$ for any $V\in\sV$, i.e., a
non-significant $V_{i'}$ is included in every
share. Therefore, we call such a non-significant share $V_{i'}$ a {\em
common} share. 
\begin{rem}\label{vac-ramp.rem}
It is known that for any access structure with $L+1$ levels, the coding
 rate $\rho_i$ must satisfy $\rho_i\ge 1/L$ for any significant
 share $V_i$ \cite{KOSOT-ecrypt93}. Especially, in the case of
 $(k,L,n)$-threshold SSSs, the optimal ramp SSS attaining
 $\rho_i=1/L$ for all $i$ can easily be constructed
 \cite{BM-crypto85,HY-ieice}. Any common share $V_i$ must
 also satisfy that $\rho_i\ge 1/L$. On the other hand, in the same
 way as Remark \ref{vac.rem} for the perfect SSSs, each vacuous share $V_i$
 can be realized as $\rho_i=0$ for any access structure. Furthermore, if  
 there exists a vacuous share with  $\rho_i>0$, the average coding rate
 can be reduced by setting $\rho_i=0$ without changing all the significant
 and the common shares. \QED
\end{rem}

Letting $\check\as_j \DEF \bigcup_{\ell=j}^L\as_\ell$ and 
$\hat{\as}_j\DEF\bigcup_{\ell=1}^j\as_\ell$, for $j=
0,1,\ldots,L$, the monotonicity in (\ref{mono1.eq}) and (\ref{mono2.eq})
are extended as follows:
\begin{eqnarray}
\label{mono-a.eq}
\sA\in\check{\as}_j~ \Rightarrow~ \sA' \in \check{\as}_j 
~\mbox{\rm for~all}~\sA' \supseteq \sA\\
\label{mono-b.eq}
\sA\in\hat{\as}_j~ \Rightarrow~ \sA' \in \hat{\as}_j 
~\mbox{\rm for~all}~\sA' \subseteq \sA
\end{eqnarray}
Therefore, the minimal and maximal families of the access structure,
$\AS^{R-}=\{\as^-_0,\as^-_1,\ldots,\as^-_L\}$ and  
$\AS^{R+}=\{\as^+_0,\as^+_1,\ldots,\as^+_L\}$, respectively, can be
defined as
\begin{eqnarray}\label{condition}
\as^-_j&=&\{\sA\in\as_j:\sA-\{V\} \not\in\check{\as}_j~\mbox{\rm for~any}
~V\in\sA\},\\
\as^+_j&=&\{\sA\in\as_j:\sA\cup\{V\} \not\in\hat{\as}_j~\mbox{\rm for~any}
~V\in\VV- \sA\}.
\end{eqnarray}
Then, the following theorem holds. 
\begin{thm}[\cite{KOSOT-ecrypt93}]\label{KOSOT.thm}
A ramp SSS with access structure $\AS^R=\{\as_0,\as_1,\ldots,\as_L\}$
 can be constructed if and only if $\check{\as}_j$ (or $\hat{\as}_j$) 
satisfies the monotonicity  (\ref{mono-a.eq}) (or (\ref{mono-b.eq})) for
 all $j=1,2,\ldots,L$. \QED
\end{thm}

In Theorem \ref{KOSOT.thm}, the necessity of the condition is obvious,
 and the sufficiency is established by the next construction. 
\begin{const}[\cite{KOSOT-ecrypt93}]\label{Kurosawa-ramp.const} 
Let 
$S=\{S^{\langle 1 \rangle},S^{\langle 2 \rangle},\ldots,S^{\langle
L \rangle}\}$ be a secret, and let $\AS^{{\langle j \rangle}}=
\{\check{\as}_j,\VV-\check{\as}_j\}$, $j=1,2,\ldots,L$, be
 the perfect access structures determined from a given access structure
 $\AS^R$. Since each $\AS^{\langle j \rangle}$ is a perfect access
 structure satisfying the monotonicity (\ref{mono1.eq}) and
 (\ref{mono2.eq}), we can construct a SSS with  $\AS^{\langle j \rangle}$ 
for secret $S^{\langle j \rangle}$. Letting $\{V^{\langle j 
\rangle}_1,V^{\langle j \rangle}_2,\ldots, V^{\langle j \rangle}_n\}$ 
be the shares for $S^{\langle j \rangle}$ and 
 $\AS^{\langle j \rangle}$, the share $V_i=\{
V_i^{\langle 1 \rangle},V_i^{\langle 2 \rangle},\ldots,V_i^{\langle
 L \rangle}\}$ realizes the access structure $\AS^R$. 
For $\AS^R$, a ramp SSS can also be
 constructed from $\{2^{\mbox{\boldmath\scriptsize$V$}}-
\hat{\as}_j,\hat{\as}_j\}$ instead of
 $\AS^{\langle j \rangle}=
\{2^{\mbox{\boldmath\scriptsize$V$}}-\check{\as}_j,\check{\as}_j\}$. 

\QED
\end{const}
\begin{rem}
Note that in Construction \ref{Kurosawa-ramp.const}, we have
 $\rho_i \ge 1$ for any access structure. For example, in the case that Construction
 \ref{Kurosawa-ramp.const} is applied to the $(k,L,n)$-threshold access
structure, the constructed ramp SSS has $\rho_i=1$ although the
 $(k,L,n)$-threshold SSS can be realized with
 $\rho_i= 1/L$. Therefore,
 Construction \ref{Kurosawa-ramp.const} is not efficient generally. 
\QED 
\end{rem}
\begin{ex}\label{K-et.al.ex}
Consider the following ramp access structure $\Gamma_4^R$ for
 $\sV=\{V_1,V_2,V_3,V_4\}$:
\begin{eqnarray}
\label{K-et.al.eq1}
\as_3&=&\{\{V_1,V_2,V_3,V_4\}\},\\
\as_2&=&\{\{V_1,V_2,V_3\},\{V_1,V_3,V_4\}\},\\
\as_1&=&\{\{V_1,V_2,V_4\},\{V_2,V_3,V_4\}\},\\
\label{K-et.al.eq4}
\as_0&=&\{\sA: 0 \le |\sA| \le 2\}.
\end{eqnarray}
First, we derive the
 access structures $\AS^{\langle 1 \rangle}$, $\AS^{\langle 2  
\rangle}$, and $\AS^{\langle 3 \rangle}$ based on
 (\ref{K-et.al.eq1})--(\ref{K-et.al.eq4}), and it is easy to see that
 $\AS^{\langle 1 \rangle}$ and  $\AS^{\langle 3 \rangle}$ become $(3,4)$-
 and $(4,4)$-threshold access structures, respectively. Hence, we have
 $V_i^{\langle 1 \rangle}=W_i^{(3)}$ and 
 $V_i^{\langle 3 \rangle}=W_i^{(4)}$ for $i=1,2,3,4$ where 
 $\{W^{(3)}_i\}_{i=1}^4$ and
 $\{W^{(4)}_i\}_{i=1}^4$ are the share sets of $(3,4)$-
 and $(4,4)$-threshold access structures for secrets 
$S^{\langle 1 \rangle}$ and $S^{\langle 3 \rangle}$, respectively. 
Furthermore, a perfect SSS with the access structure $\AS^{\langle 2 
\rangle}$ for a  secret $S^{\langle 2 \rangle}$ can be realized by 
$\{V^{\langle 2 \rangle}_i\}_{i=1}^4$ such that $V^{\langle 2 \rangle}_1=W'^{(3)}_1$, 
$V^{\langle 2 \rangle}_2=W'^{(3)}_2$, 
$V^{\langle 2 \rangle}_3=W'^{(3)}_3$, and 
$V^{\langle 2 \rangle}_4=W'^{(3)}_2$ where $\{W'^{(3)}_i\}_{i=1}^3$ is
 the share sets of $(3,3)$-threshold SSS for $S^{\langle 2 \rangle}$. 

According to Construction \ref{Kurosawa-ramp.const}, we can 
obtain the shares such that 
$V_1=\{W_1^{(3)},W'^{(3)}_1,W_1^{(4)}\}$, 
$V_2=\{W_2^{(3)},W'^{(3)}_2,$\\ 
$W_2^{(4)}\}$, 
$V_3=\{W_3^{(3)},W'^{(3)}_3,W_3^{(4)}\}$, 
$V_4=\{W_4^{(3)},W'^{(3)}_4,W_4^{(4)}\}$. Since each share consists of
 three primitive shares for three secrets $S^{\langle 1 \rangle}$,
 $S^{\langle 2  \rangle}$, $S^{\langle 3 \rangle}$, the constructed ramp
 SSS has $\tilde{\rho}=\rho^*=1$. \QED
\end{ex}

The construction of ramp SSSs for general access structures are treated
in \cite{SRR-icrypt02}. But, since the construction in
\cite{SRR-icrypt02} is based on monotone span programming, it is much
complicated compared with the multiple assignment map. 
\subsection{Optimal Multiple Assignment Maps for Ramp SSSs}\label{OMAMR.sec}
First, let 
$\sW_{(t,L,m)}=\{W_1^{(t,L)},W_2^{(t,L)},\ldots,W_m^{(t,L)}\}$ 
be the set of primitive shares for the $(t,L,m)$-threshold ramp SSS with
the coding rate $\rho_i=1/L$. Then, defining $\v{y}$ and
$\v{a}(\ell;\sA)$ in the same way as the perfect SSSs
in Section \ref{main.sec},
the optimal ramp SSS by the multiple assignment map for a general access
structure $\AS^R$ can be obtained by solving the following integer
programming problem:

\begin{center}
\begin{tabular}{rr@{}l@{}lclc}
\underline{IP${}^R_{\tilde\rho}\left(\AS^R\right)$}&&& \\
minimize    & $\vh\cdot\v{y}^T$~ && \\
 subject to & ${\v{a}(-1;\sA)}\cdot\v{y}^T$ ~&$\ge$&~ $0$  & for~&
 $\sA\in{\as}^-_L$ &\\
&$-{\v{a}(-1;\sA)}\cdot\v{y}^T$~~&$=$ &~ $j$ & for &$\sA\in{\as}^+_j \cup{\as}^-_j$ & 
for~ $1 \le j \le L-1$~~~~~~~~~~$(\star)$\\
&$ -{\v{a}(-1;\sA)}\cdot\v{y}^T$ ~&$\ge$&~ $ L$ & for& $\sA\in{\as}^+_0$ &\\
&$ \v{y}$                         ~&$\ge$&~ \bf 0 &&
\\
\end{tabular}
\end{center}

\begin{rem}\label{x_n.rem}
From the monotonicity defined in (\ref{mono-a.eq}) and (\ref{mono-b.eq}),
it is sufficient to consider only $\sA\in\as_j^+\cup\as_j^-$ instead of
all $\sA\in\as_j$ on the marked line $(\star)$ in 
IP${}^R_{\tilde\rho}\left(\AS^R\right)$. Note that the same primitive
 shares may be distributed to all shares since there may exist common
 shares in ramp SSSs. Hence, we may have $x_N \neq 0$ in the ramp
 SSSs although we can always assume that $x_N=0$ in the perfect SSSs. 
 \QED
\end{rem}

From Remark \ref{vac-ramp.rem}, significant or common shares $V_i$
must satisfy that $|\varphi_\AS(i)|\ge 1$ for any multiple assignment
map $\varphi_\AS$. On the other hand, $|\tilde\varphi_\AS(i')|=0$ must
hold for vacuous shares $V_{i'}$ for the optimal multiple assignment map
$\tilde\varphi_\AS$ attaining the minimal average coding rate. Hence, it
suffices to consider only significant shares and common shares in the
ramp SSSs. 
\begin{ex}
If the access structures $\AS^R_4$ in Example \ref{K-et.al.ex} is
 applied to the integer programming problem {\rm
 IP}${}^R_{\tilde\rho}\left(\AS^R_4\right)$, the following multiple
 assignment map is obtained 
\begin{eqnarray}
V_1 &=& \tilde{\varphi}_{\AS_4^R}(1)=\left\{W_1^{(7,3)},W_2^{(7,3)}\right\},\\
V_2 &=& \tilde{\varphi}_{\AS_3^R}(2)=\left\{W_3^{(7,3)},W_4^{(7,3)}\right\},\\
V_3 &=& \tilde{\varphi}_{\AS_4^R}(3)=\left\{W_5^{(7,3)},W_6^{(7,3)}\right\},\\
V_4 &=& \tilde{\varphi}_{\AS_4^R}(4)=\left\{W_3^{(7,3)},W_7^{(7,3)}\right\},
\end{eqnarray}
where $W_i^{(7,3)}\in\sW_{(7,3,7)}$. $\tilde{\varphi}_{\AS_4^R}$ attains that
 $\tilde{\rho}=\rho^*=2/3$. \QED
\end{ex}
Note that the coding rates less than $1$ cannot be achieved by
Construction \ref{Kurosawa-ramp.const}. Furthermore, our construction is
much simpler compared with the method in \cite{SRR-icrypt02}. But,
unfortunately, the integer programming problem may not have any feasible
solutions in the case of ramp SSSs. 
\begin{ex}
The following access structure $\AS_5^R$
 cannot be constructed by
 any multiple assignment map since the corresponding integer
 programming problem has no feasible solution.
\begin{eqnarray}
\label{OkK1.eq3}
\as^-_4 &=& \{\{V_1,V_2,V_3,V_4\},\{V_1,V_2,V_4,V_5\},\{V_2,V_3,V_4,V_5\}\},\\
\nonumber
\as  _3 &=& \{\{V_1,V_2,V_3,V_5\},\{V_1,V_3,V_4,V_5\},\{V_1,V_2,V_3\},\{V_1,V_2,V_4\},\{V_1,V_3,V_4\},\\
        & &  ~~ \{V_1,V_3,V_5\},\{V_2,V_3,V_4\}\},\\
\as  _2 &=& \{\{V_1,V_2,V_5\},\{V_1,V_4,V_5\},\{V_2,V_3,V_5\},\{V_2,V_4,V_5\},\{V_3,V_4,V_5\},\{V_1,V_3\},\{V_1,V_5\}\},\\
\as  _1 &=& \{\{V_1,V_2\},\{V_2,V_3\},\{V_3,V_4\}\},\\
\label{OkK2.eq3}
\as^+_0 &=& \{\{V_1,V_4\},\{V_2,V_5\},\{V_3,V_5\}\},
\end{eqnarray}
\QED
\end{ex}
In this case, we can modify the definition of the ramp SSS given by
(\ref{ramp-def.eq}) as follows.
\begin{eqnarray}
\label{relax1}
H(S|\sA) &=&   0,~~~~~~~~~~~~~~~\hspace*{.5mm}\mbox{\rm for~all~} \sA\in\as_L,\\
H(S|\sA) &\ge& \frac{L-j}{L}H(S),~~\hspace*{.5mm}\mbox{\rm for~all~} \sA\in\as_j,~1 \le j \le L-1,
\\
\label{relax3}
H(S|\sA) &=&   H(S),~~~~~~~~~~\mbox{\rm for~all}~ \sA\in\as_0.
\end{eqnarray}
In order to implement (\ref{relax1})--(\ref{relax3}) in the integer
programming, it suffices to replace the marked line $(\star)$ in
IP${}^R_{\tilde\rho}\left(\AS^R\right)$ by
$-{\v{a}(-1;\sA_j)}\cdot\v{y}^T \ge
j$. Letting IP${}^{R2}_{\tilde\rho}\left(\AS^R\right)$ be the modified
integer programming problem, the next theorem holds. 
\begin{thm}\label{feasible.thm}
The integer programming problem {\rm IP}${}^{R2}_{\tilde\rho}\left(\AS^R\right)$
 always has a feasible solution for any access structure $\AS^R$. \QED
\end{thm}
\noindent
{\em Proof of Theorem \ref{feasible.thm}:} Let $\cal V$ be a multiset in
$\VV$, some elements of which may be the same. Then, for $\cal V$ and
$\sA\subseteq{\sV}$, we define $N({\cal V},\sA)$ as follows.
\begin{eqnarray}
N({\cal V},\sA)=\left|\{\sA'\in{\cal V}:\sA\subseteq\sA'\}\right|,
\end{eqnarray}
where all $\sA'\in{\cal V}$ are treated as different sets even if some of
them are the same. Now we construct a multiset $\cal U$ for 
$\AS^R=\{\as_0,\as_1,\ldots,\as_L\}$ by the next construction. 
\begin{const}\label{ramp-ex.const}~
\begin{itemize}
\item[(1)] Let ${\cal U}:=\emptyset $ and $j:=1$. 
\item[(2)] For each $\sA \in \as_{L-j}^+$ satisfying $N({\cal U},\sA) < j$, we
	   add $\sA$ into ${\cal U}$, $(j-N({\cal U},\sA))$ times. 
\item[(3)] Let $j:=j+1$. 
\item[(4)] If $j<L$, go to (2). In case of $j=L$, go to (5).
\item[(5)] Output ${\cal U}$. \QED 
\end{itemize}
\end{const}
From the monotonicity of $\check{\as}_j$ in (\ref{mono-a.eq}), the 
family ${\cal U}$ can always be constructed. Then, letting ${\cal U}
=\{\sF_1,\sF_2,\ldots,\sF_m\}$, we can define a map
$\check{\psi}:\{1,2,\ldots,n\}\rightarrow
2^{\mbox{\boldmath\scriptsize$W$}_{(m,L,m)}}$ by 
\begin{eqnarray}\label{ramp-cum.eq}
\check{\psi}(i)=\bigcup_{j:V_i\not\in
{\mbox{\boldmath\scriptsize$F$}}_j}
\left\{ W^{(m,L)}_j \right\},
\end{eqnarray}
where $W^{(m,L)}_j \in \sW_{(m,L,m)}$. 
Note that in the case of $L=1$, (\ref{ramp-cum.eq}) coincides with the
cumulative map in (\ref{cum.eq}). Furthermore, for any set $\sF_\ell
\in{\cal U}$, 
we can check from (\ref{ramp-cum.eq}) that 
\begin{eqnarray}\label{prop-cum.eq}
W_{\ell'}^{(m,L)}\not\in\bigcup_{i:V_i\in{\mbox{\boldmath\scriptsize$F$}}_\ell}
\check{\psi}(i),
\end{eqnarray}
holds for all $\ell'$ satisfying $\sF_{\ell} \subseteq \sF_{\ell'}$. 

Now, assume that $\sF_{\ell}\in{\cal A}_j^+$. Then, from Construction
\ref{ramp-ex.const}, 
there exist a family of $j$ subsets $\{\sF_{\ell_1},\sF_{\ell_2},
\ldots,\sF_{\ell_j}\}\subseteq {\cal U}$ satisfying
$\sF_\ell\subseteq\sF_{\ell'}$ for 
$\ell'\in\{\ell_1,\ell_2,\ldots,\ell_{j}\}$. Hence, it holds from
(\ref{prop-cum.eq}) that $W^{(m,L)}_{\ell'} \notin
\bigcup_{i:V_i\in{\mbox{\boldmath\scriptsize$F$}}_\ell}
\check{\psi}(i)$ for $\ell'\in\{\ell_1,\ell_2,\ldots,\ell_{j}\}$. This
means that we can verify that 
$\left|\bigcup_{i:V_i\in{\mbox{\boldmath\scriptsize$F$}}_{\ell}}
\check{\psi}(j)\right|\le m-j$, and $V_i=\check{\psi}(i)$
satisfies (\ref{relax1})--(\ref{relax3}). Therefore, 
IP${}^{R2}_{\tilde\rho}\left(\AS^R\right)$ always has at least one
feasible solution. \QED

\halflineskip
Note that as shown in the following example, Construction
\ref{ramp-ex.const} gives inefficient assignments of the primitive
shares, generally. 
\begin{ex}
Assume that the access structure $\AS_5^R$ in
 (\ref{OkK1.eq3})--(\ref{OkK2.eq3}) satisfies the conditions
 (\ref{relax1})--(\ref{relax3}). First, we apply Construction
 \ref{ramp-ex.const} to the access structure $\AS_5^R$. Then, we obtain
 the following multiset ${\cal U}_{\AS_5^R}$.
\begin{eqnarray}
\nonumber
{\cal U}_{\AS_5^R} &=& \{\{V_1,V_2,V_3,V_5\},\{V_1,V_3,V_4,V_5\},\{V_1,V_2,V_4\},\{V_1,V_2,V_5\},\{V_1,V_4,V_5\},\{V_2,V_3,V_5\},\\
 & &   \{V_2,V_3,V_4\},\{V_2,V_4,V_5\},\{V_2,V_4,V_5\},\{V_3,V_4,V_5\},\{V_1,V_4\}\}.
\end{eqnarray}
Hence, we can obtain $V_i=\check{\psi}(i)$, $i=1,2,\ldots,5$, as follows:
\begin{eqnarray}
\label{cum-r1.eq}
V_1 &=& \check\psi(1) = \left\{W_6^{(11,4)},W_7^{(11,4)},W_8^{(11,4)},W_9^{(11,4)},W_{10}^{(11,4)}\right\},\\
V_2 &=& \check\psi(2) = \left\{W_2^{(11,4)},W_5^{(11,4)},W_{10}^{(11,4)},W_{11}^{(11,4)}\right\},\\
V_3 &=& \check\psi(3) = \left\{W_3^{(11,4)},W_4^{(11,4)},W_5^{(11,4)},W_8^{(11,4)},W_9^{(11,4)},W_{11}^{(11,4)}\right\},\\
V_4 &=& \check\psi(4) = \left\{W_1^{(11,4)},W_4^{(11,4)},W_6^{(11,4)}\right\},\\
\label{cum-r2.eq}
 V_5 &=& \check\psi(5) = \left\{W_3^{(11,4)},W_7^{(11,4)},W_{11}^{(11,4)}\right\},
\end{eqnarray}
where $W_i\in \sW_{(11,4,11)}$. In this case, we have 
 $\tilde\rho=21/20$ and $\rho^*=3/2$ since it holds
 that $H(W_i^{(11,4)})=H(S)/4$ for each $i$. 

On the other hand, we can construct the following optimal 
 multiple assignment map $\tilde{\varphi}_{\AS_5^R}$ by solving the integer
 programming problem ${\rm IP}_{\tilde\rho}^{R2}(\AS_5^R)$. 
\begin{eqnarray}
\label{mas-r1.eq}
V_1 &=& \tilde{\varphi}_{\AS_5^R}(1) = 
\left\{W_1^{(8,4)},W_2^{(8,4)}\right\},\\
V_2 &=& \tilde{\varphi}_{\AS_5^R}(2) = 
\left\{W_3^{(8,4)},W_4^{(8,4)},W_5^{(8,4)}\right\},\\
V_3 &=& \tilde{\varphi}_{\AS_5^R}(3) = 
\left\{W_2^{(8,4)},W_6^{(8,4)}\right\},\\
V_4 &=& \tilde{\varphi}_{\AS_5^R}(4) = 
\left\{W_7^{(8,4)},W_8^{(8,4)}\right\},\\
\label{mas-r2.eq}
V_5 &=& \tilde{\varphi}_{\AS_5^R}(5) = \left\{W_9^{(8,4)}\right\},
\end{eqnarray}
where $W^{(8,4)}_i \in \sW_{(8,4,9)}$, and it holds that
 $\tilde\rho=1/2$ and $\rho^*=3/4$, which are more efficient than the
 rates of Construction \ref{ramp-ex.const}. Note that
(\ref{cum-r1.eq})--(\ref{cum-r2.eq}) and
 (\ref{mas-r1.eq})--(\ref{mas-r2.eq}) do not satisfy (\ref{ramp-def.eq})
 but satisfy (\ref{relax1})--(\ref{relax3}). For instance, in
 (\ref{mas-r1.eq})--(\ref{mas-r2.eq}), it holds for 
 $\{V_1,V_5\}\in\as_2$ that $H(S|\{V_1,V_5\})=H(S) > H(S)/2$. 

Finally, we compare Construction \ref{Kurosawa-ramp.const} with
 Construction \ref{ramp-ex.const} for the access
 structure $\AS^R_5$. If we use the cumulative map to realize each
 perfect SSS with the access structure $\AS^{\langle j \rangle}_5$, 
 $j=1,2,3,4$, in Construction \ref{Kurosawa-ramp.const}, we obtain
 $\tilde\rho=9/5$ and $\rho^*=2$. Hence, Construction
 \ref{Kurosawa-ramp.const} is more inefficient than Construction 
\ref{ramp-ex.const} in this case. \QED
\end{ex}

\section{Conclusion}

We proposed a method to construct SSSs for any given general access
structures based on $(t,m)$-threshold SSSs and integer programming. The
proposed method can attain the {\it optimal} average and/or worst coding
rates in the sense of multiple assignment maps. Hence, the proposed
method can attain smaller coding rates compared with the cumulative maps
and the modified cumulative maps. Furthermore, the proposed method can
be applied to incomplete and/or ramp access structures in addition to
complete and perfect access structures.

\end{document}